\documentclass[twocolumn,a4paper,10pt]{article}
\usepackage{amsmath,amsfonts}
\usepackage{epsfig}
\usepackage{times}
\usepackage{epsfig}
\usepackage[small,hang]{caption2}
\makeatletter

\newcounter{numbersec}
\renewcommand{\section}[1]{\par\noindent\stepcounter{numbersec}
\par
\vspace{6pt}
\noindent\textbf{\large   \arabic{numbersec} \hspace*{0.3cm} #1 }
\par
\vspace{2pt}
}
\renewcommand{\subsection}[1]{
\par
\vspace{6pt}
\noindent\textbf{#1}
\par
}
\renewcommand{\subsubsection}[1]{%
\par
\vspace{6pt}
\textbf{#1.}
}

\newcommand{\Abstract}{\par\vspace{6pt}\noindent\textbf{\large Abstract}\par\vspace{2pt}}
\newcommand{\Acknowledgments}{\par\vspace{6pt}\noindent\textbf{\large Acknowledgments }\par\vspace{2pt}}

\newenvironment{References}{
\par\vspace{6pt}\noindent\textbf{\large References}\par\vspace{2pt}
\begin{small}\begin{list}{ }{
\itemsep0mm \parsep0mm\labelsep0mm\leftmargin0mm
}}
{\end{list}\end{small}}

\makeatother

\title{\vspace*{-12mm}
\LARGE \sc \textbf{  
Towards adaptive simulations of turbulent wings at high Reynolds numbers
}}
\author{ \large \bf \textit{ 
F. Mallor$^{1,2*}$, \'A. Tanarro$^{1,2}$, N. Offermans$^{1,2}$, A. Peplinski$^{1,2}$, R. Vinuesa$^{1,2}$, P. Schlatter$^{1,2}$} \\ 
\normalsize \bf  \textit{$^{1}$SimEx/FLOW, Engineering Mechanics, KTH Royal Institute of Technology, Stockholm, Sweden} \\
\normalsize \bf  \textit{$^{2}$Swedish e-Science Research Centre (SeRC), Stockholm, Sweden} \\ \\
\underline{\normalsize \bf $^{*}$mallor@kth.se}
}
\date{}

\begin{document}
%


%

\maketitle
\thispagestyle{empty}


%
%
\Abstract
Adaptive mesh refinement (AMR) in the high-order spectral-element method code Nek5000 is demonstrated and validated with well-resolved large-eddy simulations (LES) of the flow past a wing profile. In the present work, the flow around a NACA 4412 profile at a chord-based Reynolds number $Re_c=200,000$ is studied at two different angles of attack: 5 and 11 degrees. The mesh is evolved from a very coarse initial mesh by means of volume-weighted spectral error indicators, until a sufficient level of resolution is achieved at the boundary and wake regions. The non-conformal implementation of AMR allows the use of a large domain avoiding the need of a precursor RANS simulation to obtain the boundary conditions (BCs). This eliminates the effect of the steady Dirichlet BCs on the flow, which becomes a relevant source of error at higher angles of attack (specially near the trailing edge and wake regions). Furthermore, over-refinement in the far field and the associated high-aspect ratio elements are avoided, meaning less pressure-iterations of the solver and a reduced number of elements, which leads to a considerable computational cost reduction. Mean flow statistics are validated using experimental data obtained for the same profile in the Minimum Turbulence Level (MTL) wind tunnel at KTH, as well as with a previous DNS simulation, showing excellent agreement. This work constitutes an important step in the direction of studying stronger pressure gradients and higher Reynolds complex flows with the high fidelity that high-order simulations allow to achieve. Eventually, this database can be used for the development and improvement of turbulence models, in particular wall models.

%
%
\section{Introduction}

In recent years, due to the introduction of regulations affecting aviation pollutant emissions, drag reduction has become a the major focus of the aerospace industry. One of the main sources of drag comes from the development of turbulent boundary layers (TBLs) around the aircraft wings. Due to the wing curvature, these TBLs are subjected to strong adverse and favourable pressure gradients (PG), which significantly affect the boundary-layer physics and development.  These effects are still not fully understood and characterized. Traditionally, these flows have been studied experimentally (see for instance Coles and Wadcock (1979) or Wadcock (1987)). However, as pointed out by Barlow \textit{et al.} (2015), wind-tunnel testing of wing profiles is prone to suffering from test-section wall interference in the flow and uncertainty from the employed measurement techniques. 
Recent developments in the available computational resources and methods have allowed to accurately simulate the TBLs developing around wing profiles (see Sato \textit{et al.} (2016), Fr\`ere \textit{et al.} (2018), Hosseini \textit{et al.} (2016) and Vinuesa \textit{et al.} (2018) for instance). However, the Reynolds numbers achieved so far are still moderate, and the computational domains small (particularly in the spanwise direction, a fact that affects the development of the wake). 

Providing a database of well-resolved simulation results for the flow around wing profiles is relevant due to a number of reasons; In the first place, specific aspects of the pressure-gradient effects may be studied, and different scaling properties of the boundary layer may be deduced. Similarly, such a database may be employed for the development, assessment and improvement of turbulence models, in particular including wall models in the presence of non-equilibrium conditions such as high pressure gradients. This aspect is particularly important as evidenced by the recently defined canonical speed-bump geometry by Boeing and NASA (see Slocknick 2019) which is specifically designed to provide a well-defined favorable-adverse pressure gradient sequence. Finally, providing an efficient setup for wing simulations can be used as a test bed for extended studies on wings such as the effect of control and roughness, free-stream turbulence and aeroelasticity. 

In this work, we aim to increase both the achievable $Re_c$ (Reynolds number based on freestream velocity $U_{\infty}$ and chord length $c$), vary the angle of attack, and optimize the domain size by using adaptive mesh refinement (AMR). AMR was first introduced by Berger and Oliger (1984), and later adapted for general finite-element method (FEM) simulations by Johnson and Hansbo (1992). In a previous work (Tanarro et al., 2020), we validated the implementation at a moderate angle of attack of 5$^\circ$, using a previous conformal simulation by Vinuesa et al. (2018) as reference. In this study, apart from the previous AMR simulation at 5$^\circ$, we extend the achievable angle of attack to 11$^\circ$, and we validate our implementation against both DNS (Hosseini \textit{et al.} (2016)) and experimental data (Mallor \textit{et al.} (2021)). The framework of generating the initial mesh, the refinement strategy and the extraction of statistics is very general, such that this work constitutes an important initial step with the aim of extending our work to a higher $Re_c=1,640,000$, as in the experimental work by Wadcock (1987). This latter Reynolds number is then relevant for practical applications.

\section{Methodology}
\begin{figure}[htp]
	\begin{center}
		\includegraphics*[width=0.95\linewidth]{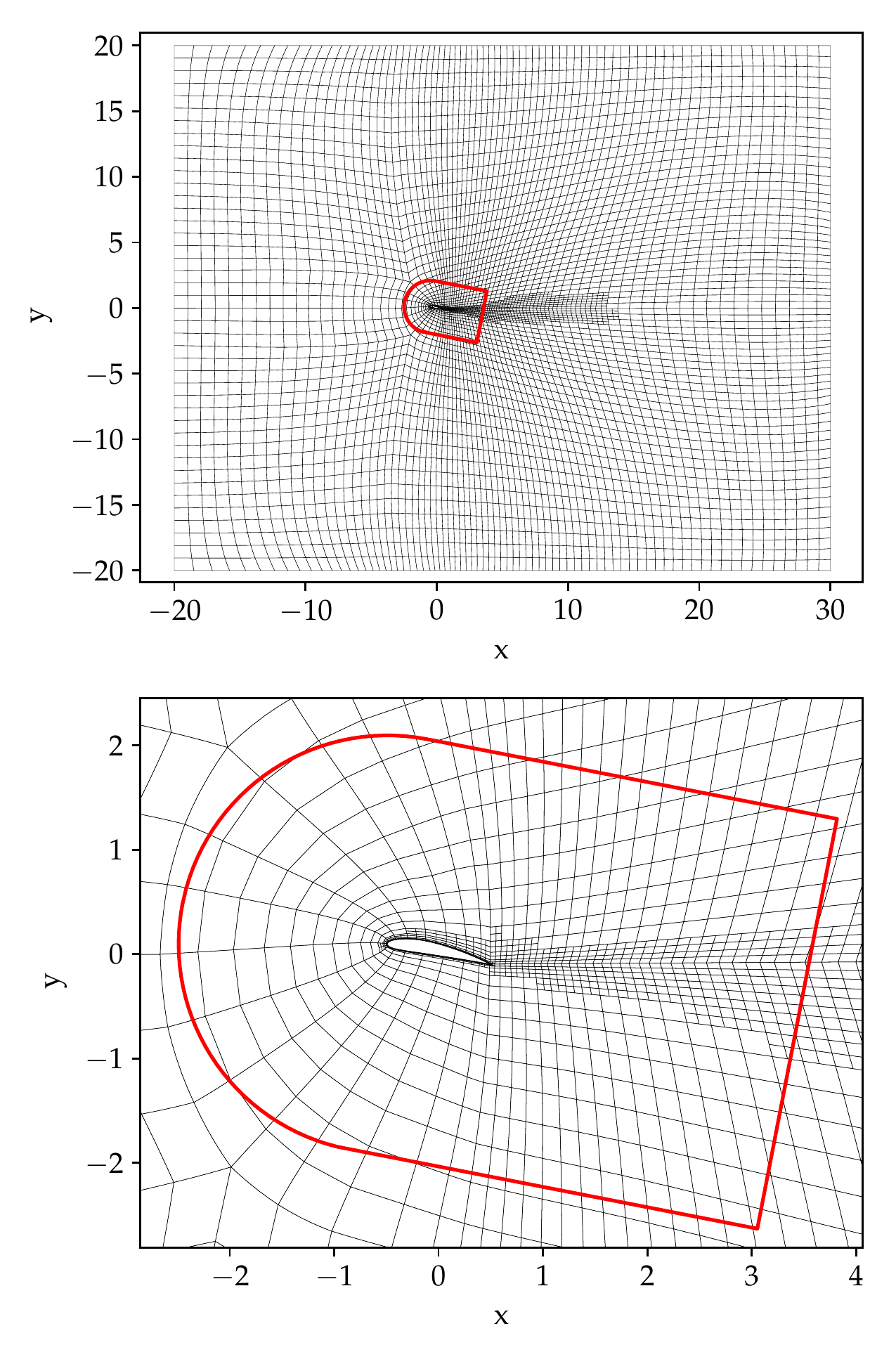} 
		\caption{\label{fig:mesh} Refined mesh (spectral-element boundaries shown in the plot, each element consists of $8\times8\times8$ collocation points) for the case with 11$^\circ$ angle of attack, with the domain of a typical conformal simulation superposed in red.}
	\end{center}
\end{figure}

The AMR setup validation is performed based on the results for the flow around a NACA 4412 wing profile at a 5$^\circ$ and 11$^\circ$ angle of attack and $Re_c=200,000$. Well-resolved large-eddy simulations (LES) using an explicit relaxation filter, analogous to the one described in Schlatter \textit{et al.} (2004) are carried out using the spectral-element  code Nek5000 (Fischer \textit{et al.}, 2008). Nek5000 is particularly noted for its efficient implementation and excellent parallelization properties, running on millions of concurrent ranks. In the work by Offermans (2019) and Offermans \textit{et al.} (2020), a non-conformal version of Nek5000 was implemented, based on the p4est library and using appropriate modified preconditioners. Compared to previous LES and DNS performed on the same wing case, the present AMR setup was completely redesigned, and presents several notable differences. Firstly, the domain size ($L_x,L_y,L_z$) is considerably larger: ($50c,40c,0.4c$ for 5$^\circ$, and $0.6c$ 11$^\circ$) in the AMR case and ($6c, 4c, 0.2c$) in previous conformal cases. Note that $x$, $y$ and $z$ denote the horizontal, vertical and spanwise directions, respectively. The difference in domain size can clearly be seen in Figure \ref{fig:mesh}, in which the typical domain boundary of a conformal simulation is superposed over the mesh of an AMR simulation. Secondly, the polynomial order used in previous simulations ($N=11$) is higher than the one of the AMR simulation ($N=7$), which gives a more uniform local resolution and thus higher stable time steps. Lastly, the boundary conditions in conformal cases have to be obtained from a precursor RANS simulation due to the proximity of the domain boundaries to the wing. As this is not an issue in the AMR case, a uniform Dirichlet boundary condition is imposed in the inlet, top and bottom boundaries (both cases have an stabilized outflow condition).

\begin{figure}[htp]
	\begin{center}
		\includegraphics*[width=0.95\linewidth]{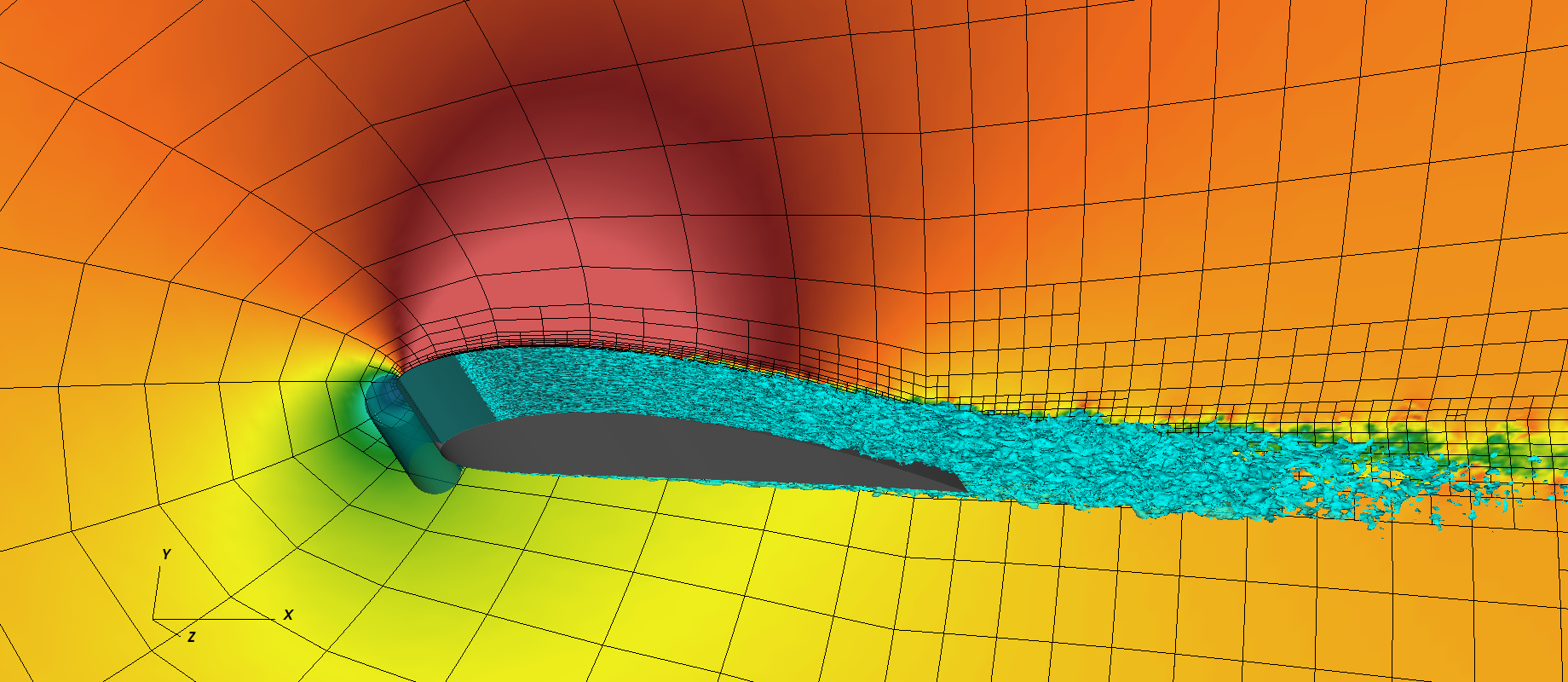} \\ \vspace{0.4cm}
		\includegraphics*[width=0.95\linewidth]{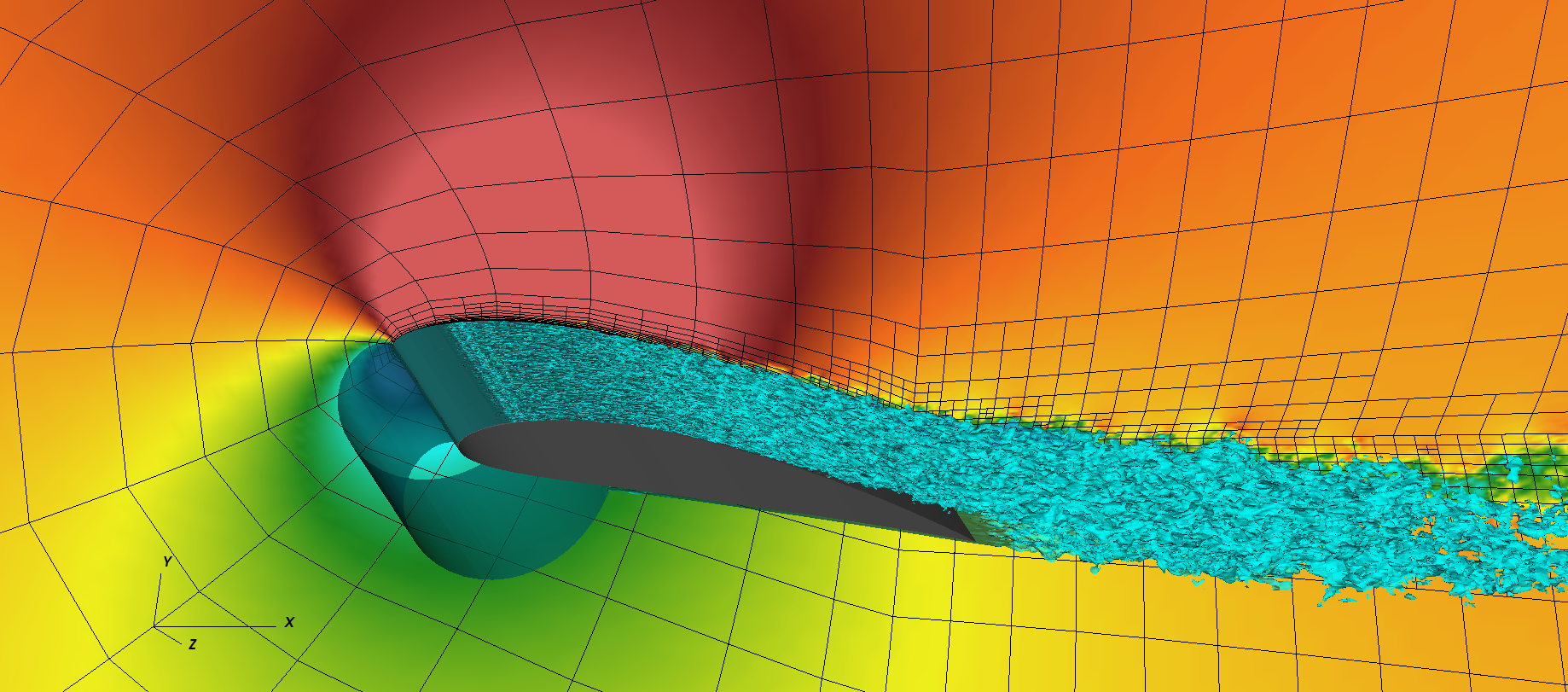}
		\caption{\label{fig:isocontour} Three-dimensional isocontours (in green) of $U_x =0.75 U_\infty$. The background shows a two-dimensional slice of the horizontal velocity, as well as a projection of the spectral element boundaries. The top and bottom figures correspond to 5$^\circ$ and 11$^\circ$ angle of attack, respectively.}
	\end{center}
\end{figure}
In order to generate the mesh, the following procedure is followed: Starting from an initial coarse mesh, refinement is performed progressively, $i.e.$ elements are refined or coarsened, until a final mesh, used for the production runs, is fixed. The so-called $h$-type refinement (element splitting) is used with a fixed frequency, based on volume-weighted spectral error indicators (developed by Mavriplis (1990) and further discussed in the context of Nek5000 by Offermans \textit{et al.}, 2020) collected in each element during each period of refinement. The resulting meshes, which are shown in Figure \ref{fig:isocontour}, concentrate the refinement in areas of larger velocity oscillations (\textit{i.e.}\ areas with turbulence), allowing to properly resolve the complex turbulent flow both in the boundary layer and wake regions. Note that error indicators are averaged in time, which yields a mesh that will not adapt to individual turbulent fluctuations, but rather give a grid that is optimized in a time-averaged sense. 

The current study also includes an experimental campaign conducted concurrently at KTH. This study is motivated by the lack of accurate boundary-layer resolved experimental data on wing surfaces, where the non-uniform pressure-gradient evolution and the inflow conditions are controlled. For the time being, the experimental data used in this work for the validation of mean flow quantities (pressure and velocity) is obtained using the wind tunnel setup described extensively in Mallor \textit{et al.} (2021). In short, a fiber-glass NACA 4412 wing profile is mounted in the test section of the Minimum Turbulence Level (MTL) wind tunnel at KTH. The wing model is equipped with 65 pressure ports located both on the suction and pressure sides of the airfoil (as well as one in the trailing edge), which allows to measure the pressure distribution along the surface. Velocity data is obtained by means of hot-wire (HW) anemometry scans (performed both in the boundary layer and wake regions). Two-point correlation (using 2 HWs) scans in the wake are used to determine the region in which the flow is correlated. A correct sizing of the spanwise domain in the AMR simulations, which allows to capture the largest structures present is achieved using this data. It should be noted that the flow around the wing is tripped (\textit{i.e.}\ transition to turbulence is forced) on both sides of the chord (at $x/c=10\%$, clearly visible in Figure \ref{fig:isocontour}) both in the experimental and numerical realizations of the flow. The tripping of the boundary layer is particularly important for cases at lower angles of attack where no leading-edge separation bubble would appear naturally (see Wadcock 1987 and Fr\`ere \textit{et al.} 2018) to ensure that turbulence is initiated at the same location and thus the flow is not dependent on unknown inflow conditions.

\section{ Results}

In this section, mean flow quantities obtained using the numerical setup described previously (and shown in Figures \ref{fig:mesh} and \ref{fig:isocontour}) are shown and validated, including pressure and velocity measurements in the wind-tunnel experiments. For the sake of conciseness, results related to turbulence fluctuations will be presented at a later stage.

\begin{figure}[htp]
	\begin{center}
		\includegraphics*[width=0.95\linewidth]{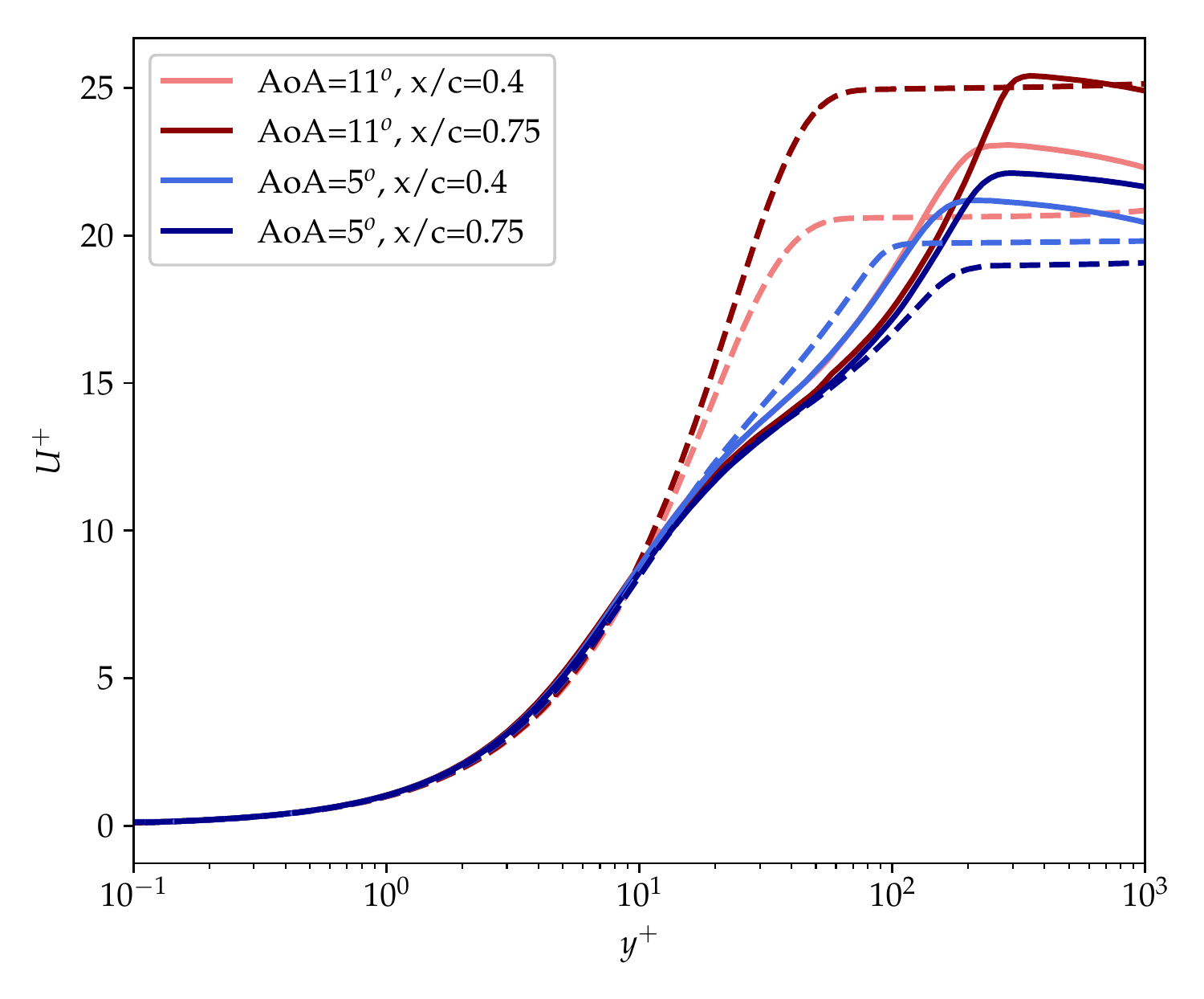} 
		\caption{\label{fig:meanU} Mean velocity profiles at $x/c=0.4$ and $0.75$ for both the 5$^\circ$ and 11$^\circ$ angle of attack cases. Solid and dashed lines correspond to the suction and pressure sides, respectively.}
	\end{center}
\end{figure}
Figure \ref{fig:meanU} reports the mean velocity profiles at two different chord-wise locations ($x/c$=0.4 and 0.75) for both angles of attack. The pressure gradient effect on the TBL is apparent at a first glance: As the adverse pressure gradient increases along the suction side, the wake region increases substantially, which is accentuated as the angle of attack is increased. On the pressure side, the opposite effect is observed: by increasing the angle of attack, the favorable pressure gradient (FPG) increases, which in turn leads to an almost re-laminarization of the TBL. This effect is only observed in the 11$^\circ$ case (whose pressure-side profiles closely resemble those of a sink flow), as the FPG in the 5$^\circ$ case is very mild, making the TBL almost a zero-pressure gradient one. Note also that for the cases with significant APG the velocity in the freestream is no constant which is related to the effect of deceleration and the choice of $y^+$ as surface normal direction. All profiles are normalized using the local friction velocity, thus a perfect agreement close to the wall is obtained.

\begin{figure}[h!]
	\begin{center}
		\includegraphics*[width=0.99\linewidth]{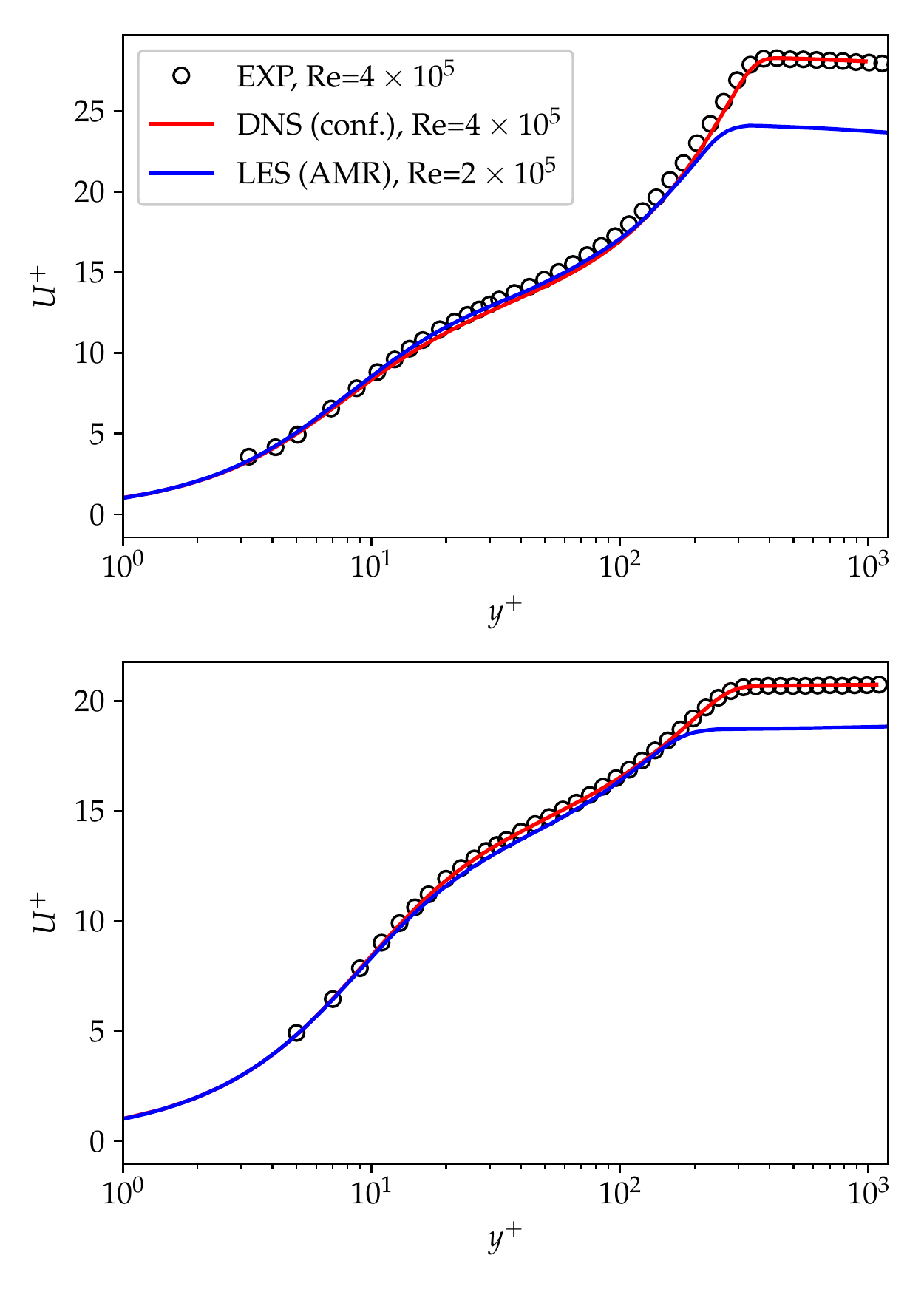} 
		\caption{\label{fig:Uvalidation} Mean velocity profiles at $x/c=0.75$ in the suction (top plot) and pressure (lower plot) sides. All data is for a 5$^\circ$ angle of attack, and the Reynolds number is of $4\times 10^5$ for the experimental and DNS data, and $2\times 10^5$ for the AMR data.}
	\end{center}
\end{figure}

In our previous work (Tanarro \textit{et al.} 2020), we performed a validation of the AMR implementation using a 5$^\circ$ case, and a previous conformal simulation (Vinuesa \textit{et al.} (2018)) as a baseline. AMR was demonstrated to lead to comparable results both in integral, mean and fluctuating (\textit{rms}) quantities while reducing the overall computational cost. In this work, we turn our attention to the experimental data recently obtained at the MTL wind tunnel. However, due to the difficulties encountered in tripping the flow at a Reynolds number of $2\times 10^5$, the only experimental data available is at $4\times 10^5$. The experimental pressure data was already validated against high-fidelity numerical data in Mallor \textit{et al.} (2021). In order to complete the validation of the experimental data, mean velocity profiles on the suction and pressure sides at a 5$^\circ$ angle of attack are reported in Figure \ref{fig:Uvalidation}. The experimental data is on top of the DNS data (Hosseini \textit{et al.} (2016)). Moreover, although the Reynolds number used is lower, the AMR data agrees quite well in the inner and logarithmic regions, with both the DNS and experimental data. The effect of $Re$ is of course visible in the wake region and the final value of $U_\infty^+$. This overall agreement is important, as the pressure gradient at that moderate angle of attack is almost constant as a function of Reynolds number (see the discussion in Vinuesa \textit{et al.} 2018), making the flow around the NACA 4412 airfoil relevant in the study of pressure and high-Re effects on TBLs.

One important aspect is the identification of the wall position and the correct friction value for the experimental data. We are currently working on extending the available composite profiles to include also the effect of pressure gradients (see Mallor \textit{et al.} 2021). For the presentation of the data here, we have identified the best fit to the DNS data to identify the surface position (shift in $y$, and the friction.

Another important mean quantity, which is highly dependent on the wake (and, therefore, the boundary layer development), is the pressure distribution around the wing surface. We report it as the pressure coefficient,
\begin{equation}
	c_p = \frac{p-p_{\infty}}{\frac{1}{2}\rho U_\infty} \ ,
\end{equation}
where ${\rho}$, $p_{\infty}$, and $U_{\infty}$ refer to the ambient density and pressure, and inflow velocity, respectively. Note that in incompressible simulations, it is not immediately clear how normalize the pressure as in principle any addititve constant is possible. Therefore, we chose to use as normalization the averaged pressure over a large stretch on the suction side of the airfoil, and in this way eliminate possible effects of outflow conditions.

\begin{figure}[htp]
	\begin{center}
		\includegraphics*[width=0.99\linewidth]{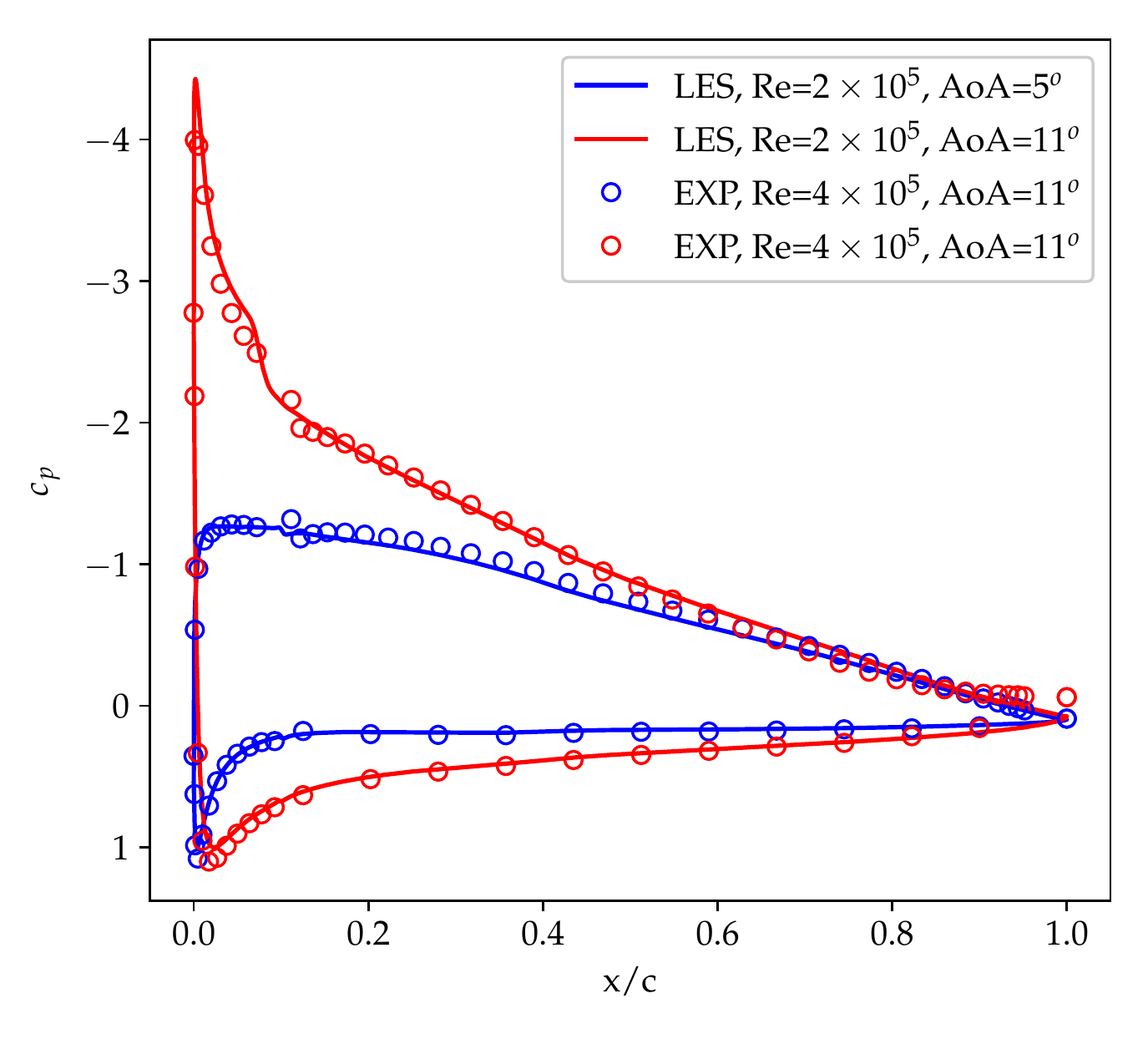} 
		\caption{\label{fig:Pvalidation} Pressure coefficient ($c_p$) distribution along the wing surface for the 11$^\circ$ (in red) and 5$^\circ$ (in blue) angle of attack cases. Numerical data from the AMR simulation is shown as a solid line, and experimental pressure data as open circles.}
	\end{center}
\end{figure}

The results reported in Figure \ref{fig:Pvalidation} show excellent agreement between the numerical (AMR) and experimental $c_p$ data. The collapse of both datasets is almost perfect for the 5$^\circ$ case, and only minor differences on the suction side near the trailing edge are present in the 11$^\circ$ case. These differences could be indicative of a slight acceleration in the test section around the suction side due to the blockage introduced by the wing when placed at a high angle of attack. This acceleration of the flow results in a slightly increased "effective" angle of attack, leading to an earlier flow separation. Also, the effect of tripping at $x/c$ may be seen in particular on the suction side. As discussed above, fixing this location is essential to ensure similar flow development and thus allow for a relevant comparison. Nonetheless, the agreement is outstanding on the pressure side and along the majority of the suction side.
Note that the agreement in pressure coefficient at both angles of attack is despite the fact of a twofold difference in Reynolds numbers; the velocity profiles and local friction are quite different (see Fig.~4). This observation strengthens the importance of the choice of the underlying wing profile.

\section{ Conclusions and outlook}
In this work, the implementation of adaptive mesh refinement and non-conformal meshing in the spectral-element method code Nek5000 is tested and validated through well-resolved LES simulations of the flow past a NACA 4412 wing profile at a chord-based Reynolds number of 200,000 and angles of attack of 5 and 11 degrees. The data acquired using AMR, which was previously validated against conformal simulations in Tanarro \textit{et al.} (2020), is further validated using an experimental dataset obtained with the same wing model in the Minimum Turbulence Level (MTL) tunnel at KTH with the setup described in Mallor \textit{et al.} (2021). Both mean velocity profiles (Fig. \ref{fig:Uvalidation}) in the turbulent boundary layer, and the pressure distribution around the airfoil (Fig. \ref{fig:Pvalidation}) are found to be in good agreement with the experimental data. Minor differences encountered near the trailing edge for the 11$^\circ$ case can be attributed to blockage effects in the test section of the experiments in the MTL. The present case is particularly tricky to simulate correctly, as the large adverse pressure gradient results in mean flow separation near the trailing edge. This has several implications in terms of required domain size: turbulent structures become much larger in the spanwise direction (we found via two-point synchronized HW measurements in the wake that the flow to be correlated up to $z/c$=0.25), causing any simulation with an insufficient spanwise domain to fail at correctly representing the flow in this regime. Moreover, pressure fluctuations become stronger and affect a larger area. The previous conformal simulations, in which the domain boundaries (in which a steady Dirichlet BC based on a precursor RANS simulation was prescribed) were only two chord lengths away from the airfoil (and therefore wake), which would lead to strong interactions with these fluctuations.
Therefore, this study, together with the one by Tanarro \textit{et al.} (2021), leads to the conclusion that by combining the advantages of AMR and high-order methods, highly complex turbulent flows, such as that of a wing profile, can be correctly modeled and studied to a high degree of precision and flexibility, while reducing the computational costs and uncertainties involved. Doing this type of simulations using a conformal structured mesh would be essentially impossible at reasonable meshing effort. 

The aim is to exploit these computational savings and improved setup, in order to increase the complexity of the turbulent flows achievable via high-order simulations. In the context of pressure-gradient TBLs this will be achieved following two different approaches. On the one hand, by increasing the Reynolds number (with the aim of ultimately achieving a $Re_c = 1.64\times 10^6$, as in the experimental work by Wadcock 1987) at a moderate angles of attack (5$^\circ$), high-Re effects on pressure-gradient (both adverse and favourable) can be studied. This follows from the fact that the pressure gradient distribution at moderate angles of attack in the NACA 4412 is almost independent of $Re$. On the other hand, the increase in angle of attack will allow to study much larger adverse pressure gradients, and their effect on the flow (such as the increase in backflow events and the onset of mean flow sepatation). Therefore, this work constitutes a basis and validation of this larger body of research.

The development of the adaptive simulation framework has been improved in various aspects, including the refinement strategy, the choice of preconditioners and load balancing. For the present case, the non-conformal mesh does not pose any additional penalty on the computation costs. Therefore, all the gains in terms of time step, accuracy and cost of meshing directly benefit the actual time to solution.


\Acknowledgments
Simulations were performed on resources provided by the Swedish National Infrastructure for Computing (SNIC) through grant agreement no. 2018-05973 at PDC Center for High Performance Computing, KTH, Stockholm. This research is funded by the Knut and Alice Wallenberg Foundation as part of the Wallenberg Academy Fellow programme, and was partially supported by the ExaFLOW H2020 Project (Project Number 671571) and the Excellerat Centre of Excellence.

\begin{References}

\item Barlow JB, Rae Jr WH, Pope A (2015) Low speed wind tunnel testing. \textit{INCAS Bulletin} 7(1):133

\item Berger MJ, Oliger J (1984) Adaptive mesh refinement for hyperbolic partial differential equations. \textit{J Comput Phys} 53:484--512

\item Coles D, Wadcock AJ (1979) Flying-hot-wire study of 
ow past an NACA 4412 airfoil at maximum lift. \textit{AIAA J} 17(4):321--329

\item Fischer P, Kruse J, Mullen J, Tufo H, Lottes J, Kerkemeier S (2008) Nek5000: Open source spectral element CFD solver. Available at: https://nek5000.mcsanl.gov/

\item Fr\`ere A, Hillewaert K, Chatelain P, Winckelmans G (2018) High Reynolds number airfoil: from wall-resolved to wall-modeled LES. \textit{Flow Turbul Combust} 101:457--476

\item Hosseini, SM, Vinuesa, R, Schlatter, P, Hanifi, A, Henningson, DS (2016) Direct numerical simulation of the flow around a wing section at moderate Reynolds number. \textit{Int. J. Heat Fluid Flow} 61:117–-128

\item Johnson C, Hansbo P (1992) Adaptive finite element methods in computational mechanics. \textit{Comp
Meth Appl Mech Engng} 101:143--181

\item Mallor, F, Parikh, A, Doga, E, Atzori, M, Hajipour, M, Tabatabaei, N, \"{O}rl\"{u}, R, Vinuesa, R, Schlatter, P (2021), KTH Internal Report

\item Mavriplis C (1990) A posteriori error estimators for adaptive spectral element techniques. In: Wesseling P (ed) \textit{Notes on Numerical Fluid Mechanics}, pp 333--342

\item Offermans N (2019) Aspects of adaptive mesh refinement in the spectral element method. \textit{PhD Thesis KTH}

\item Offermans N, Peplinski A, Marin O, Schlatter P (2020). Adaptive mesh refinement for steady flows in Nek5000. \textit{Computers and Fluids} 197, 104352

\item Sato M, Asada K, Nonomura T, Kawai S, Fujii K (2016) Large-eddy simulation of NACA 0015 airfoil flow at Reynolds number of 1.6$\times10^6$. \textit{AIAA J} 55(2):673--679

\item Schlatter P, Stolz S, Kleiser L (2004) LES of transitional flows using the approximate deconvolution model. \textit{Int J Heat Fluid Flow} 25(3):549--558

\item Slotnick, J P (2019) Integrated CFD Validation Experiments for Prediction of Turbulent
Separated Flows for Subsonic Transport Aircraft. NATO Report STO-MP-AVT-307.

\item Tanarro, A, Mallor, F, Offermans, N, Peplinski, A, Vinuesa, R, Schlatter, P (2020) Enabling adaptive mesh refinement for spectral-element simulations of turbulence around wing sections. \textit{Flow, Turbulence and Combustion}, 105:415--436.

\item Vinuesa R, Negi P, Atzori M, Haniffi A, Henningson D, Schlatter P (2018) Turbulent boundary layers around wing sections up to $Re_c$=1,000,000. \textit{Int J Heat Fluid Flow} 72:86--99

\item Wadcock AJ (1987) Investigation of low-speed turbulent separated flow around airfoils. \textit{NASA Tech Rep} (NASA-CF- 177450)

\end{References}
\end{document}